 \documentstyle[12pt,epsf,aaspp4]{article}

  \def\kpc{{\rm\,kpc}}

  \def\msun{{\rm\,M_\odot}}

  \def\vol#1  {{{#1}{\rm,}\ }}
  \def\lya{${\rm Ly}\alpha$}
  \def\etal{et al.\ }

  \newcount\refno
  \newcount\rfno
  \def\eq{$^{\the\refno\ }$\advance\refno by 1}
  \def\ad{\advance\rfno by 1}

  \def\clock{\count0=\time \divide\count0 by 60
  \count1=\count0 \multiply\count1 by -60
 \advance\count1 by \time
           
 \number\count0:\ifnum\count1<10{0\number\count1}\else\number\count1\fi}

\def\Gcm2{\rm G~cm^2}

\def\beq{\begin{equation}}
\def\eeq{\end{equation}}

\def \date         {\ifcase\month \message{zero} \or
                    January \or February \or March \or
				  April \or May \or June 
                   \or July \or 
                         August \or September \or October
		  \or November \or 
									                      December \fi
                \space\number\day, \number\year}

\begin{document}
\title{Detection and Fundamental Applications of Individual First Galaxies}
\author{Renyue Cen\altaffilmark{1}}
\altaffiltext{1} {Princeton University Observatory,
		  Princeton University, Princeton, NJ 08544;
		  cen@astro.princeton.edu}
\received{\date}
\accepted{ }

\begin{abstract}

First galaxies formed within halos of mass $M=10^{7.5}-10^9\msun$ at $z=30-40$ 
in the standard cold dark matter (CDM) universe 
may each display an extended hydrogen 21-cm absorption halo 
against the cosmic microwave background 
with a brightness temperature decrement of $\delta T=-(100-150)$~mK 
at a radius $0.3 \le r\le 3.0$~comoving Mpc,
corresponding to an angular size of $10-100$ arcseconds.
A 21-cm tomographic survey in the redshift shell $z=30-40$
(at $35-45$MHz),
which could be carried out by the next generation of radio telescopes,
is expected to be able to detect millions of first galaxies and 
may prove exceedingly profitable in enabling (at least) four 
fundamental applications for cosmology and galaxy formation.
First, it may yield direct information on star formation physics in first galaxies.
Second, it could provide a unique and sensitive probe
of small-scale power in the standard cosmological model 
hence physics of dark matter and inflation.
Third, it would allow for an independent,
perhaps ``cleaner" characterization of interesting features
on large scales 
in the power spectrum such as the baryonic oscillations.
Finally, possibly the most secure,
each 21-cm absorption halo
is expected to be highly spherical 
and faithfully follow the Hubble flow.
By applying the Alcock-Paczy\'nski test
to a significant sample of first galaxies,
one may be able to determine the dark energy equation of state 
with an accuracy likely only limited
by the accuracy with which the matter density
can be  determined independently.

\end{abstract}

\keywords{galaxies - radio - intergalactic medium - cosmology: theory}

\section{Introduction}

It is of wide interest to detect and understand the first generation of galaxies,
expected to form in the redshift range $z=30-50$ 
in the standard CDM universe (Spergel \etal 2003).
Extensive literatures on 21-cm properties of neutral hydrogen in the dark ages  
and during cosmological reionization 
have long focused on large-scale fluctuations of the intergalactic neutral hydrogen  
and global spectral features (e.g., Hogan \& Rees 1979; Scott \& Rees 1990).
In this {\it Letter} we point out a unique
feature possessed by the first {\it individual} galaxies of mass
$10^{7.5}-10^9\msun$ formed at $z=30-40$ --- a large hydrogen 21-cm absorption halo
against the cosmic microwave background (CMB).
Each 21-cm absorption halo has a size 
$10^{''}-100^{''}$ with a brightness temperature decrement of $\delta T=-(100-150)$~mK
at $35-45$MHz,
which could serve as a visible proxy for each galaxy that otherwise may 
be undetectable. 
The next generation of radio telescopes, such as LOFAR,
may be able to detect such a signal.
A range of fundamental applications is potentially possible with 
a redshift (i.e., 21-cm tomographic) survey of the first galaxies
in the redshift shell $z=30-40$,
which may hold the promise to revolutionize the field of cosmology
and shed illuminating light on dark matter, dark energy and inflation physics.
Throughout, a standard (Wilkinson Microwave Anisotropy Probe) 
WMAP-normalized CDM model 
is used (unless indicated otherwise): $\Omega_M=0.31$, $\Lambda=0.69$,
$\Omega_b=0.048$, $H_0=69$km/s/Mpc, $n_s=0.99$ and $\sigma_8=0.90$.

\section{Large 21-cm Absorption Halos of First Galaxies}

A first-generation galaxy is expected to emit UV and X-ray radiation,
each carving out an H II region of size (ignoring recombination): 
$r_{\hbox{HII}} \sim 43 ({M_h\over 10^7\msun})^{1/3} ({c_*\over 0.1})^{1/3} ({f_{esc}\over 0.1})^{1/3} ({N_p\over 8\times 10^4})^{1/3}\kpc$ comoving,
where $M_h$ is the halo mass,
$c_*$ the star formation efficiency, 
$f_{esc}$ the ionizing photon escape fraction into the intergalactic medium (IGM),
and $N_p$ the number of hydrogen ionizing photons 
produced by each baryon formed into stars,
($\sim 10^{4.5-5}$ for a massive metal-free Population III IMF;
Bromm, Kudritzki, \& Loeb 2001).
Hard X-ray photons ($\ge 1$keV) produced escape deep
into the IGM with a distance of $\sim 500-1000$ comoving megaparsecs,
building an X-ray background.
Sandwiched between small H II regions and the X-ray sea
sits a quite large \lya\ scattering region (Loeb \& Rybicki 1999),
resulting in a four-layer structure, as depicted in Figure 1.

\begin{figure}
\plotone{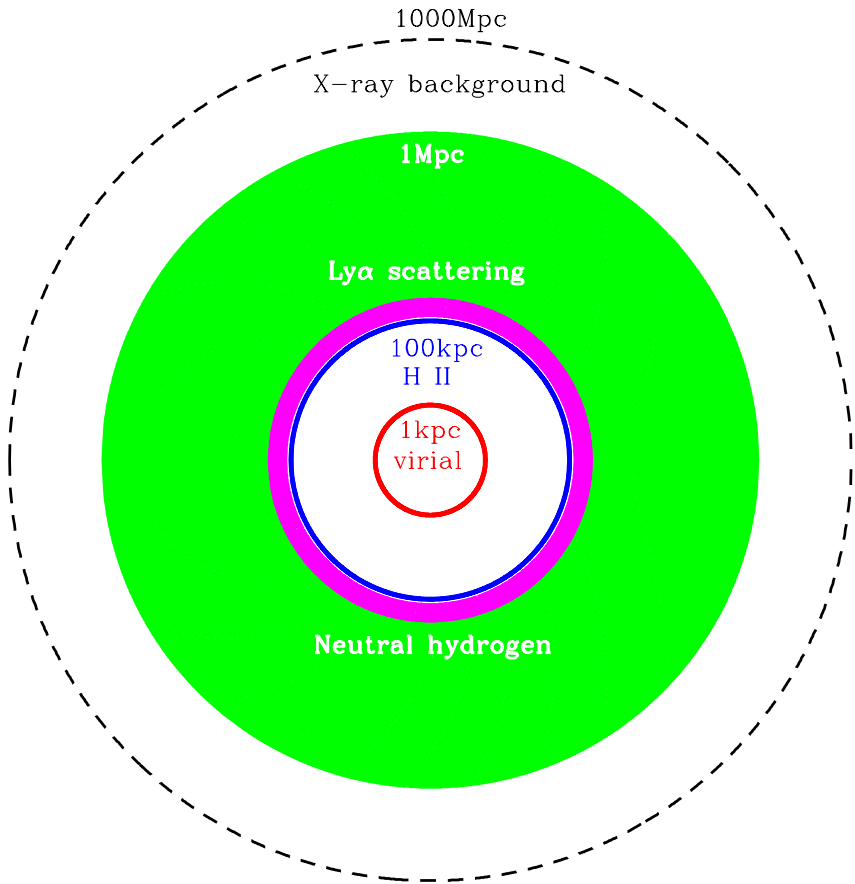}
\vskip -2cm
\caption{
shows a four-layer structure around an individual
galaxy at very high redshift.
The inner-most central region (inside the red circle)
is the virialized region where star formation occurs with a typical
size of about $1~$kpc comoving.
The next region, enclosed by the blue circle, is the ionized
region with a typical size of order $40-200$kpc comoving.
Exterior to the ionized region lies the \lya\ scattering region
shown in magenta for the inner (heated) part and in green for the outer (cold) part,
where \lya\ photons can strongly couple 
the 21-cm spin temperature to the kinetic temperature of the gas;
the inner part shown in magenta is significantly heated by UV and soft X-ray
photons emanating from the host galaxy, while
most of the region shown in green remains cold at a temperature
set by the general IGM.
Outside the green circle is the general IGM that is only 
affected collectively by cumulative X-ray background (as well
as a \lya\ background), as symbolically
circumscribed by the dashed black circle.
}
\label{fig1}
\end{figure}

The IGM in the vicinity 
of a galaxy interacts with ionizing UV and soft X-ray  
as well as near \lya\ photons emanating from the galaxy, which can be computed.
The most important and relevant physical processes 
are (1) the interaction between
neutral gas and near \lya\ photons emitted by the central host galaxy,
which couples the spin temperature of the IGM to its kinetic temperature
(Wouthuysen 1952; Field 1958),
and (2) the interaction between neutral gas and  ionizing 
UV and soft X-ray photons emanating from the host galaxy,
which provides a heating source for the otherwise cold IGM
up to some small radius.
In the absence of heating the kinetic temperature $T_k$ of the IGM 
would be equal to $T_{\hbox{IGM}} \approx 18 \left({1+z\over 31}\right)^2~\hbox{K}$
(for the redshift range of interest here) since
beginning of decoupling with CMB at $z\sim 200$ (Peebles 1993).
We perform spherically the transfer of UV and soft X-ray radiation from 
the host galaxy outward to compute
(1) the development of the HII region around the galaxy as a function of time,
(2) the evolution of the temperature of the surrounding IGM,
subject to UV and soft X-ray heating by the host galaxy,
as a function of radius and time,
and (3) the \lya\ coupling coefficient $y_\alpha$ as a function of radius and time.

We assume a uniform IGM density equal to
the mean gas density of the universe (i.e., $\Delta=0$ in equation 1 below).
Two scenarios of metal-free star formation in first galaxies are considered:
(1) all stars have a single mass of $200\msun$ (VMS),
advocated by Oh \etal (2001) and Qian \& Wasserburg (2002),
and 
(2) an IMF has the Salpeter slope of $2.35$ with a lower cutoff of
$25\msun$ and an upper cutoff of $120\msun$ (SAL),
close to what favored by Umeda \& Nomoto (2003), 
Tumlinson, Venkatesan, \& Shull (2004) and Tan \& McKee (2004).
We adopt the library of stellar spectra and ages from Schaerer (2002);
we use a black-body radiation spectrum for each star of chosen mass
with an effective surface temperature from Table 3 of Schaerer (2002)
but slightly adjusted so as to 
produce the correct ratio of the number of photons above helium II Lyman limit
to the number of photons above hydrogen Lyman limit,
averaged over the lifetime of each star (see Table 4 of Schaerer 2002).
We use an escape fraction for Lyman limit photons $f_{esc}$ from the host galaxy
and self-consistently a frequency dependent
escape fraction for other UV and soft X-ray photons assuming that
they are subject to the same absorbing column in the galaxy.
All escaped photons (emerging from the virial radius) 
are then subject to the (time-dependent) combined absorption
of H I, He I and He II in the IGM,
self-consistently computed.
Since the \lya\ scattering region is mostly neutral
with a residual ionized fraction of $2\times 10^{-4}$ left from
recombination (Peebles 1993),
we assume that $\eta=14\%$ of X-ray energy is used to heat the gas
(Shull \& Van Steenberg 1985) with the heating
rate per hydrogen atom at radius $r$ computed with the following formula:
${dE\over dt} = \int_0^\infty {\eta L_\nu\over 4\pi h\nu r^2}(\sigma_\nu(HI)(h\nu-h\nu_{H}) + \xi\sigma_\nu(HeI)(h\nu-h\nu_{HeI}) 
+\xi\sigma_\nu(HeII)(h\nu-h\nu_{HeII})) e^{-\tau_\nu} d\nu$,
where 
$L_\nu$ is the luminosity per unit frequency of the galaxy;
$\nu_{HI}$, $\nu_{HeI}$ and $\nu_{HeII}$ are
ionization potentials
of H, He I and He II, respectively;
$\sigma_\nu(HI)$
$\sigma_\nu(HeI)$ and 
$\sigma_\nu(HeII)$ are 
photo-ionization cross-sections 
of H, He I and He II, respectively;
$\xi$ is ratio of helium number density to hydrogen number density;
$\tau_\nu$ is the optical depth from the galaxy 
to radius $r$ at frequency $\nu$.

The spin temperature of neutral hydrogen (Field 1958, 1959) is then given by 
$T_{\hbox{s}}={T_{\hbox{cmb}}+y_\alpha T_k + y_c T_k\over 1+y_\alpha+y_c}$,
where $y_c\equiv {C_{10}\over A_{10}} {T_*\over T_k}$ 
is the collisional coupling coefficient with 
the collisional de-excitation rate 
$C_{10}={4\over 4} \kappa (1-0) n_H$,
$\kappa (1-0)$ taken from Zygelman (2005),
$n_{\rm H}$ is the mean hydrogen density
and 
$T_*=0.0682$~K is the hydrogen hyperfine energy splitting.
In the expression for the \lya\ coupling coefficient
$y_\alpha = {P_{10}T_*\over A_{10}T_k}$, 
$A_{10}=2.87\times 10^{-15}$~s$^{-1}$
is spontaneous emission coefficient of the 21-cm line, 
the indirect de-excitation rate $P_{10}$ of the hyperfine structure
levels is related to the total \lya\  scattering rate
$P_\alpha$ by $P_{10}=4P_\alpha/27$ (Field 1958).
Here $P_\alpha=\int F_\nu \sigma(\nu) d\nu$
with $F_\nu$ being the \lya\ photon flux (in units of cm$^{-2}$~s$^{-1}$)
and $\sigma(\nu)=\sigma_\alpha \phi(\nu) = {3\over 8\pi}\lambda_\alpha^2 A_\alpha \phi(\nu)$
being the cross section for \lya\ scattering (MMR),
where $\lambda_\alpha=1.216\times 10^{-5}$~cm 
is the wavelength of the \lya\ line,
$A_\alpha=6.25\times 10^8$~s$^{-1}$ is the spontaneous Einstein coefficient
for \lya\ line and $\phi(\nu)$ is the normalized \lya\ line (Voigt) profile with $\int \phi(\nu) d\nu=1$.
Most of the \lya\ scattering is accomplished 
by UV photons slightly on the blue side of the \lya\ 
that redshift into \lya\ resonant line due to the Hubble expansion
(note that $\Delta\nu/\nu\sim 10^{-3}$ due to Hubble expansion 
at $r\sim 1$Mpc comoving),
not the intrinsic \lya\ line photons that escape from the host 
galaxy and redshift to the damping wing
(Madau, Meiksin, \& Rees 1997; MMR).
Additional physical processes that
were not treated previously, including higher-order
Lyman lines that result in cascade in two-photon emission,
fine structure of \lya\ resonance and spin-flip scattering,
introduce corrections of order unity
to $P_\alpha$ (CM; Hirata 2005; Chuzhoy \& Shapiro 2005)
but all these corrections terms 
are insignificant for our case,
and we only apply the relatively large 
correction term $S_c$ ($\sim 1.5$) as shown in Figure 4 of CM
due to a spectral shape change near \lya\ .
The observed brightness temperature increment/decrement against the CMB is 
\begin{eqnarray}
\delta T = 41 (1+\Delta)x_H ({T_{\hbox{s}}-T_{\hbox{cmb}}\over T_{\hbox{s}}})({\Omega_b h^2\over 0.02})({0.15\over \Omega_Mh^2})^{1/2} ({1+z\over 31})^{1/2}~\hbox{mK},
\end{eqnarray}
\noindent
where $\Delta$ is gas overdensity relative to the mean,
$x_H$ neutral hydrogen fraction,
$T_{\hbox{cmb}}=2.73(1+z)$~K CMB temperature
and other symbols have their usual meanings.

\begin{figure}
\plotone{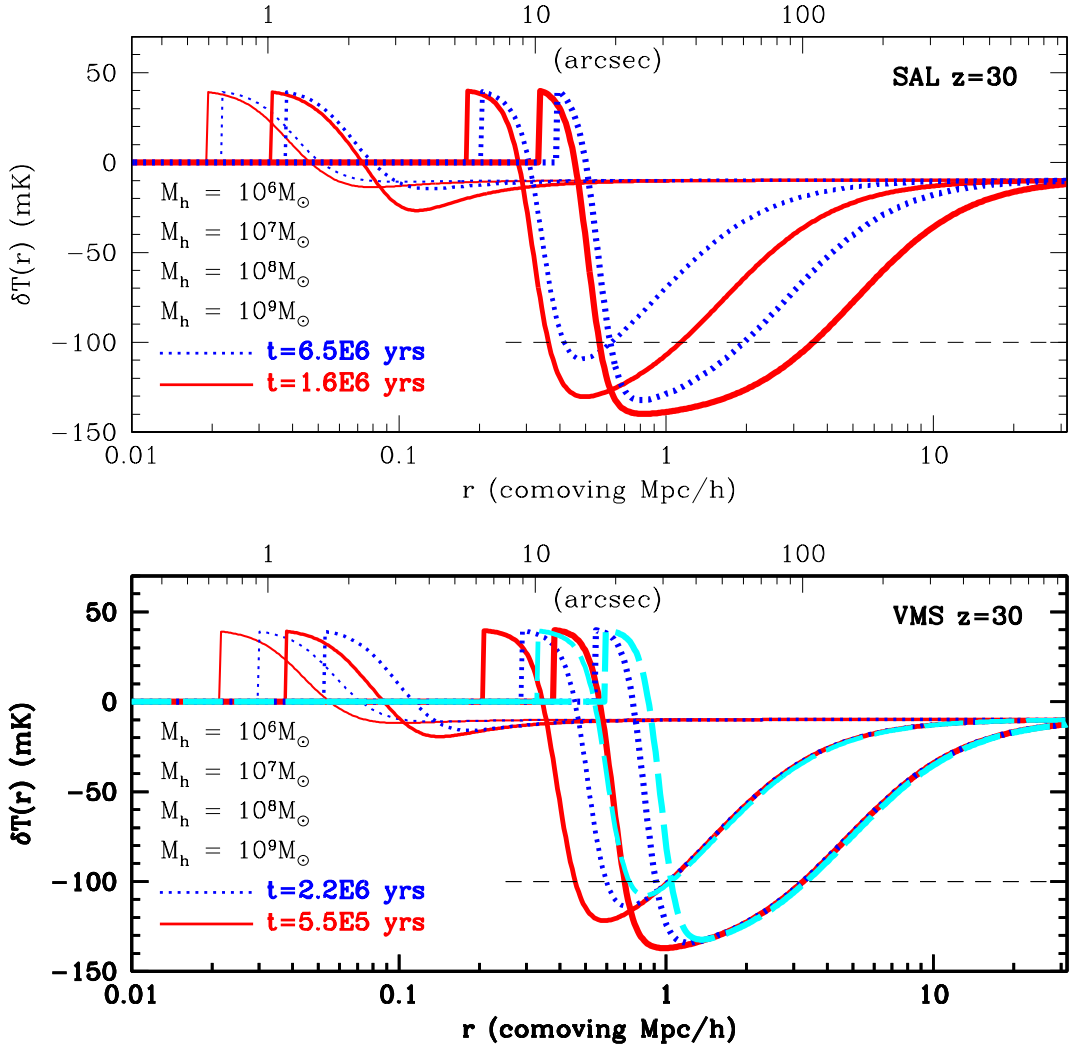}
\vskip -0.9cm
\caption{
The top panel shows the $\delta T$ profiles 
for halos with $M_h=(10^6,10^7,10^8,10^9)\msun$, respectively,
from top to bottom, at $z=30$ as a function of comoving radius,
with the adopted SAL IMF (SAL, see text).
The top x-axis is in units of arcseconds.
The division mass between large halos and minihalos at
$z=30$ is $\sim 1.4\times 10^7\msun$.
For each chosen value of $M_h$ two snapshots are shown,
at $t=1.6\times 10^6$yrs (solid curves) and $t=6.5\times 10^6$yrs (dotted curves),
where $t=6.5\times 10^6$yrs is the lifetime of the 
least massive stars with $M=25\msun$ in the chosen SAL IMF.
The bottom panel is similar to the top panel but for
the adopted VMS IMF, and the two snapshots are at 
$t=5.5\times 10^5$yrs (solid curves) and $t=2.2\times 10^6$yrs (dotted curves),
where $t=2.2\times 10^6$yrs is the lifetime of the 
stars with $M=200\msun$.
Also shown as two long-dashed curves in the bottom panel 
are two cases for $M_h=(10^8,10^9)\msun$, respectively,
with soft X-ray ($h\nu>100$eV) intensity artificially boosted by 
a factor of $10$,
which should be compared to the respective cases shown
as blue curves located slightly to their left.
The star formation efficiency is assumed to be 
$0.10$ for large halos
and $0.001$ for minihalos (Abel \etal 2002).
Escape fraction $f_{esc}=0.01$ is used, although the results 
depend very weakly on it.
}
\label{fig1}
\end{figure}

Figure 2 shows the profile of 
$\delta T$ for four cases of halo masses with each choice of IMF.
Let us examine each of the four regions (sketched in Figure 1)
with respect to 21-cm observations.
Inside the virial radius (the red circle in Figure 1)
the gas is overdense with $\delta\ge 100$ and 
a positive large-amplitude emission signal may result, 
if a significant amount of neutral hydrogen gas exists within.
However, the size of this regions falls below $0.1^"$ 
and its signal is unlikely to be detectable in the foreseeable future.
The H II region (inside the blue circle in Figure 1) 
is ionized hence $\delta T=0$.
In the region outside the \lya\ scattering region (exterior to
the green region in Figure 1)
the spin temperature of the IGM has been progressively 
attracted to the temperature of the CMB with gradually weakening 
coupling to the gas kinetic temperature by atomic collisions,
producing a small but non-negligible 21-cm absorption signal
at the redshift of interest ($z\sim 30-40$) (Loeb \& Zaldarriaga 2004).

It is the \lya\ scattering region 
that is of most interest here.
In the inner part of the \lya\ scattering region
(shown in magenta in Figure 1)
the IGM is significantly heated by UV and soft X-ray photons 
to exceed the CMB temperature, while
its spin temperature is very strongly coupled to its kinetic temperature
by \lya\ scattering.
As a result, the inner radial region at 
$0.01-0.04$~Mpc/h comoving for minihalos and $0.04-0.4$~Mpc/h 
comoving for large halos
displays an emission signal against CMB with an amplitude of $\delta T\sim 30$~mK
(a shark dorsal fin-like feature in Figure 2).
Going outward (shown in green in Figure 1),
the soft X-ray heating abates (because the
cumulative optical depth to these photons increases quickly) but 
the \lya\ scattering remains strong,
up to a distance of about $10$~Mpc/h comoving.
Consequently, a strong 21-cm absorption signal against the CMB 
with an amplitude of $\delta T= -(100-150)$~mK at $\sim 35-45$MHz (for $z=30-40$)
on a scale of $0.3-3$~Mpc/h comoving, corresponding to an angular scale
of $10^{''}-100^{''}$, is produced for large halos.
This is the 21-cm absorption halo  --- a unique and strong feature for the 
large first galaxies.
We note that the absorption signal cast by minihalos 
(the two top sets of thin curves in each panel 
in Figure 2) is relatively weak
due to a combination of low mass and low star formation efficiency.
We will therefore focus on large halos for practical observability purposes.

Besides stars,
no other soft X-ray source in the galaxy is assumed to concurrently exist.
We shall examine the validity of this assumption in detail.
The density of the interstellar
medium plays an important role for
some of the potentially relevant processes considered here
and it is assumed to be $n(z) = n_0 (1+z)^3$, 
where local interstellar density $n_0=1$~cm$^{-3}$.
This assumption should hold in  
hierarchical structure formation model for the following reasons.
The mean gas density scales as $(1+z)^3$ and 
halos at low and high redshift in cosmological
simulations show similarities when
density and length are measured in their respective comoving units
(e.g., Navarro, Frenk, \& White 1997; Del Popolo 2001).
The spin parameters
(i.e., angular momentum distribution) of both high and low
redshift halos have very similar
distributions 
peaking at a nearly identical value $\lambda\sim 0.05$
(Peebles 1969; White 1984; 
Barnes \& Efstathiou 1987; 
Ueda \etal 1994;
Steinmetz \& Bartelmann 1995;
Cole \& Lacey 1996; 
Bullock \etal 2001).
Thus, cooling gas in galaxies at low and high redshift
should collapse by a similar factor 
before the structure becomes dynamically stable
(e.g., rotation support sets in), 
resulting in interstellar densities scaling as $(1+z)^3$.
Direct simulations (Abel \etal 2002; Bromm \etal 2002)
suggest a gas density of $10^3-10^4$cm$^{-3}$ by
the end of the initial free fall for minihalos at $z\sim 20$,
verifying this simple analysis.

We will estimate each of several possible
types of soft X-ray emission sources in turn. 
First, let us estimate soft X-ray emission from supernova remnants.
Assuming the standard cooling curve (Sutherland \& Dopita 1993)
we find that at $z=30$
a supernova blastwave with an initial explosion energy of $5\times 10^{52}$~erg
(for a star of mass $200\msun$; e.g., Heger \& Woosley 2002)
would enter its rapid cooling phase at a temperature of $2.7\times 10^7$~K.
This implies that the energy emitted at $\sim 100-300$~eV 
from the cooling shell is about $7\%$.
A $200\msun$ mass would release $2.5\times 10^{54}$~erg total energy
due to nuclear burning, out of which $0.3\%$ is released
in photons at $100-300$eV for our adopted radiation spectrum.
Therefore, the ratio of total photon energy from the supernova remnant 
to that from the star is $0.4-0.5$.
Thus, for the VMS IMF, stellar soft X-ray appears to dominate
over that from its supernova remnant.
The soft X-ray contribution from supernova remnant cooling  
increases relatively compared to that
from the star itself with decreasing stellar mass and we estimate that
the overall contribution from the two components may become comparable
for the SAL IMF case, averaged over time.
Second, we will examine X-rays produced  
from cooling of supernova-accelerated relativistic electrons by CMB photons
via inverse Compton (IC) process (e.g., Oh 2001).
For adiabatic shocks, as is appropriate in our case,
the IC spectral energy distribution 
has a two-power-law form:
$L_\nu\propto$~constant at $E<E_{break}$
and 
$L_\nu\propto \nu^{-1}$ at $E>E_{break}$.
The break energy is 
$E_{break}=70$~keV, independent of redshift
with the assumed scaling of the interstellar medium density with redshift.
Then the ratio of energy from IC to that from stars
is found to be $10^{-4}-10^{-3}$ in the $100-300$eV band,
depending on the exact upper energy cutoff (assuming
$10\%$ of supernova explosion energy is utilized to accelerate
relativistic electrons in shocks).
Clearly, contribution to soft X-rays from IC process is unimportant.
Third, X-ray binaries during the relatively short lifetime of massive 
stars may be rare, for top-heavy IMFs of concern here.
We can make an estimate based on
the calculation by Rappaport, Podsiadlowski \& Pfahl (2005),
who give an ultra luminous X-ray binary formation 
rate of $3\times 10^{-5}$ per supernova.
It is clear that even if each X-ray binary is able to release 
as much energy as in a supernova explosion itself and all in the 
soft X-ray band, the resulting contribution will be less than
a fraction of a percent of that from stars.
Fourth, stellar mass black holes (BH) of $\sim 10-100\msun$
may be produced in significant numbers with a top-heavy IMF
as well as a central galactic BH.
It seems that stellar BH accretion is likely 
significantly suppressed and small due to 
feedback effect from stars on surrounding gas
(e.g., Mori, Umemura, \& Ferrara 2004;
Alvarez, Bromm, \& Shapiro 2005).
A concomitant contribution of soft X-rays from central BH accretion
in the lifetime of a $200\msun$ star is approximately
$(M_{BH}/M_*)\times (1/0.007)) \times (t_*/t_E) \times (f_{BH,SX}/f_{*,SX})
=0.6 f_{BH,SX} (M_{BH}/M_*/0.003)$,
after inserting soft X-ray emission fraction for a $200\msun$ stellar spectrum
of $f_{*,SX}=0.003$, stellar lifetime $t_*=2.2\times 10^6$~yrs
and Eddington time $t_E=4.4\times 10^8$~yrs,
where $f_{BH,SX}$ is the energy fraction
released by the BH accretion in the soft X-ray band ($100-300$eV).
Thus, if the ratio of black hole mass to (bulge) stellar mass 
follows the local Magorrian (Magorrian \etal 1998) relation, 
then, unless most of the accretion energy
is released in the soft X-ray band, contribution from central BH accretion 
to the soft X-ray band is relatively small.
Fifth, thermal bremsstrahlung emission from gravitational
shock heated gas is likely negligible due to a low
gas temperature ($T\sim 10^4$K).
Finally, soft X-rays from massive first stars themselves are thought to be 
produced by stellar winds and quite uncertain.
Recent work suggests that the winds hence soft X-ray emission 
from metal-free stars are expected to be insignificant (e.g., Krticka \& Kubat 2005).
In summary, soft X-rays from neglected, possible 
sources other than that from the stellar photospheres
would, at most, make a modest correction to what is adopted in our calculation.
To ascertain our conclusion,
we compute a case with the amplitude of soft X-ray intensity
at $h\nu \ge 100$~eV artificially raised by a factor of $10$
and do not find any significant effect that would qualitatively
change our results (Figure 2).
The reason is that the IGM quickly becomes optically thick to 
a few $100$eV soft X-ray photons at $\sim 1$~Mpc comoving.
Therefore, our results should be quite robust.

\begin{figure}
\plotone{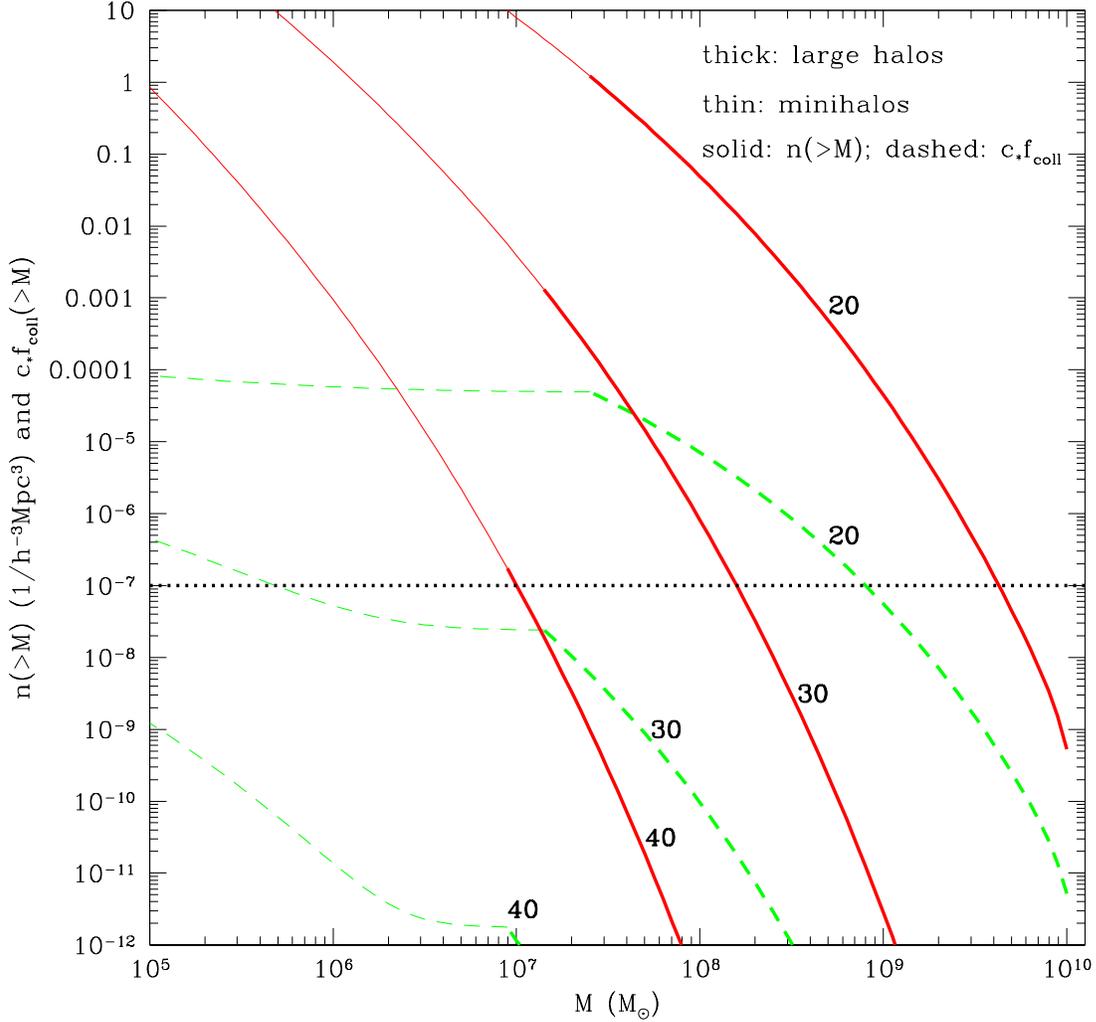}
\vskip -0.5cm
\caption{
shows cumulative halo mass functions (solid curves)
at redshift $z=(20,30,40)$, respectively, from top to bottom.
Each solid curve is broken into two parts with
a thick portion corresponding to large halos with efficient atomic
cooling
and 
a thin portion corresponding to minihalos with molecular cooling only.
The corresponding dashed curves are cumulative $c_*f_{\hbox{coll}}$ 
for the three cases, assuming star formation efficiency
of $c_*(\hbox{large})=0.1$ for large halos and 
of $c_*(\hbox{mini})=0.001$ for minihalos (Abel \etal 2002).
The horizontal dotted line indicates $c_*f_{\hbox{coll}}=10^{-7}$ (see equation 2). 
}
\label{fig1}
\end{figure}

Since we are concerned  with gas of relatively low temperature $\sim 20$K,
heating by a cumulative (hard) X-ray background may become relevant at some redshift.
We estimate when this may happen in the CDM model.
While an X-ray background may be generated by a variety of processes,
black hole accretion at the centers of galaxies
are thought to be the most dominant (e.g., Ricotti \& Ostriker 2003; 
Kuhlen \& Madau 2005),
estimated as follows.
Let us suppose the energy extraction efficiency from black
accretion is $\alpha$ and a fraction $f_x$ of the released energy
is in the form of hard X-rays. 
Then, the X-rays may collectively 
heat up the IGM temperature at most (ignoring Compton cooling and
assuming all X-ray photons in the background are consumed by the IGM) by an increment
\begin{equation}
\Delta T_{\hbox{xray}} = 1.1({f_{\hbox{coll}}\over 10^{-6}}) ({c_*\over 0.1})({\alpha\over 0.1}) ({M_{BH}/M_*\over 0.003}) ({f_x\over 0.029})~\hbox{K}
\end{equation}
\noindent
(assuming 14\% of X-ray energy is used to heat the IGM; Shull \& Van Steenberg 1985),
where $f_{\hbox{coll}}$ is the fraction of matter that 
has collapsed to halos where stars have formed.
Under the reasonable assumption that the parameters 
have their adopted fiducial values
(for $\alpha$ see Yu \& Tremaine 2002, 
$f_x$ see Elvis \etal 1994,
for $c_*$ see Gnedin 2000),
it becomes evident that,
for the very first galaxies formed in the universe
that comprise a collapsed  
mass fraction less than $10^{-6}$,
heating of the IGM by an X-ray background radiation field may be small.
Figure 3 shows cumulative halo mass functions at redshift $z=(20,30,40)$,
based on Press-Schechter (1974) formalism 
(using $\delta_c=1.67$ and the standard model of Spergel \etal 2003),
which should be accurate for the exponentially falling regime of interest here
(Sheth \& Tormen 1999; Jenkins \etal 2001).
Figure 3 suggests that at redshift $z\ge 30$ 
heating of the IGM by an X-ray background is small. 
Neglected contributions to the X-ray background
from other sources (X-ray binaries,
supernova remnants, etc) (e.g., Oh 2001, Cen 2003)
will likely just add a modest numerical
correction factor for equation (2) and the net effect, if any,
would push the epoch for significant heating by an X-ray background
to a slightly higher redshift
(note that structure formation is exponentially increasing 
with decreasing redshift at $z=30-40$).
But even a factor of a few upward correction to equation (2)
would still leave the temperature of 
IGM $z=30$ relatively unaffected by an X-ray background.

Another critical issue is whether heating of IGM by \lya\ photons is important. 
In a recent accurate calculation based on Fokker-Planck approximation,
Chen \& Miralda-Escud\'e (2004; CM) show that the heating rate by \lya\ photons 
is much lower than previous estimates (MMR).
We recast their important result (equation 17 of CM and using Figure 3 of CM),
the heating rate per hydrogen atom and per Hubble time, $\beta$,
in the following way:
\begin{equation}
\beta\equiv {\Gamma_c\over H n_{H} k_{B}} = 0.08 ({P_\alpha\over P_{th}})~\hbox{K}
\end{equation}
\noindent
at $z=30$, where
$k_{\rm B}$ is the Boltzmann constant,
$H$ is the Hubble constant
and $P_{th}=2.4\times 10^{-11}({1+z\over 31})$~s$^{-1}$
is the thermalization rate for \lya\ scattering at $z=30$,
above which \lya\ scattering brings down 
the spin temperature to the gas kinetic temperature (MMR).
The Hubble time is $1.4\times 10^8$~yrs at $z=30$.
So, over the duration of a stellar lifetime $6\times 10^6$~yrs
(of the least massive stars in our model, $25\msun$),
the gas will be heated up by $0.003$~K at
${P_\alpha\over P_{\rm th}}=1$.
For the regime of interest where we see the strong 
21-cm absorption signal (Figure 2)
we find ${P_\alpha\over P_{\rm th}}=1-10$.
Thus, heating of surrounding IGM by \lya\ photons
emanating from the host galaxy can be safely neglected.
In addition, 
since we are concerned with early times when
the universe is far from being ionized and the number of \lya\ photons
per hydrogen atom is significantly less than unity,
indicating that heating by the background \lya\ photons can also be safely neglected
(CM). Furthermore, heating rate by high order Lyman series photons
is still lower and thus negligible (Pritchard \& Furlanetto 2005).

\section{Fundamental Applications}

We have demonstrated a unique feature of first galaxies.
A large 21-cm survey of the first galaxies will be invaluable.
Demanding that each of the physical quantities 
be resolved by a factor of $10$
would translate to the following requirements:
an angular resolution of $\sim 1^{''}$,
a spectral resolution of $\sim 4$kHz
($\Delta\nu\sim 40$kHz across a radius of $1$Mpc/h at $z=30$ due to the Hubble expansion)
and a sensitivity of $\sim 10$~mK at $35-45$~MHz.
Among the next generation of radio telescopes currently under construction/consideration,
LOFAR (http://www.lofar.org) appears to be best positioned 
to be able to carry out such a survey,
at least for some of the brighter and larger galaxies.
LOFAR is currently designed to 
reach a frequency as low as $10$MHz with an angular resolution 
of $3^{''}-4^{''}$ at $35-45$MHz, a sensitivity of $10$~mK 
and a spectral resolution (i.e., processing capability)
of $1$~kHz (e.g., Rottgering 2003),
while MWA (http://web.haystack.mit.edu/arrays/MWA/index.html),
PaST (http://web.phys.cmu.edu/$\sim$past/index.html)
and SKA (http://www.skatelescope.org) appear to be placed
out of the $35-45$MHz range, as they stand now.

\begin{figure}
\plotone{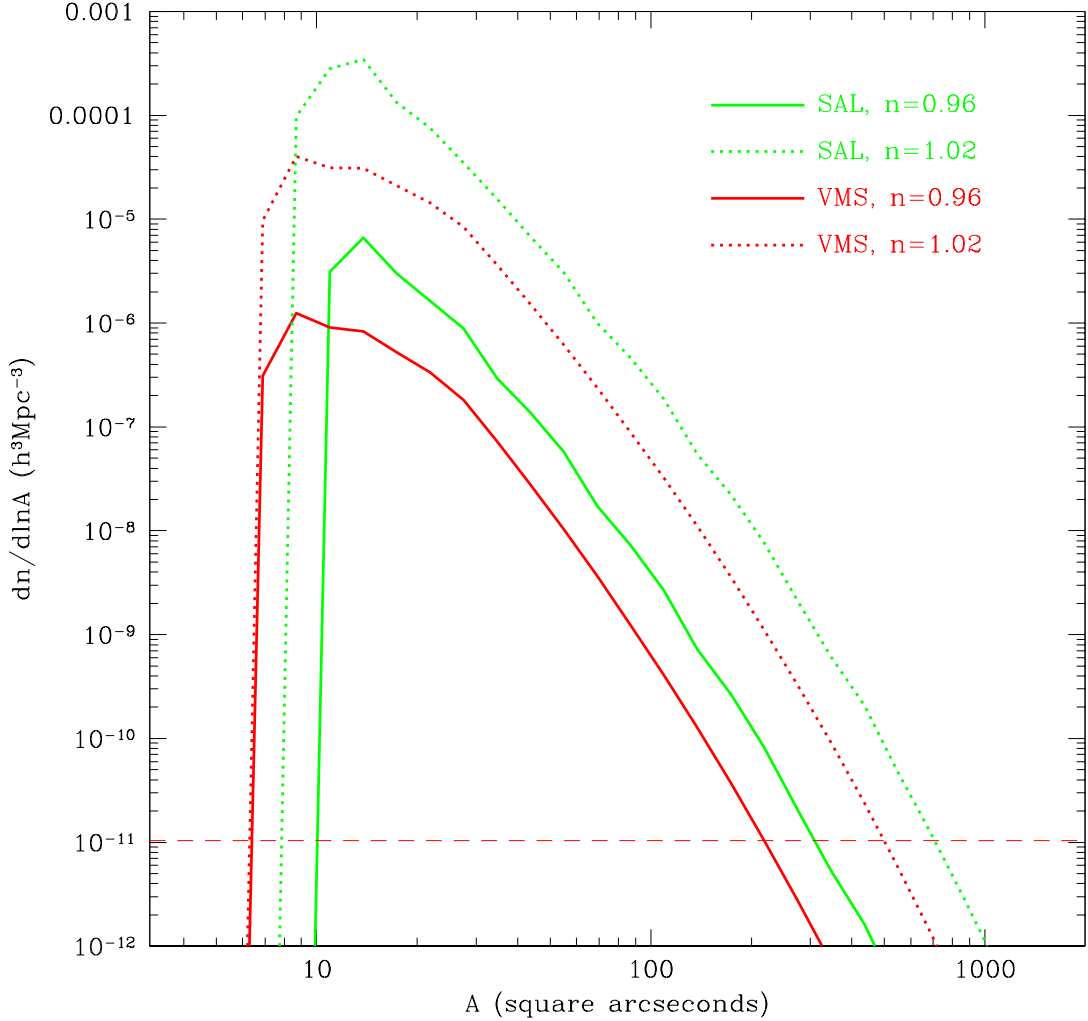}
\vskip -0.9cm
\caption{
shows the density of galaxies versus
the maximum absorption cross section 
(i.e., in the plane
of the large circle centered on the galaxy 
perpendicular to the line of sight)
with $\delta T<-100$~mK at $z=3$ in the standard LCDM model
except the index of the power spectrum $n_s$, for which 
two values are chosen, $n_s=0.96$ (solid curve) and $_s=1.02$ (dotted curve),
possibly bracketing  its likely range.
Two cases for two IMF are shown with the thick curves
for SAL and the thin curves for VMS.
The density is computed taking into account the time variations
of stellar radiation spectra and finite lifetime.
The horizontal dashed curve indicates the density with which
the volume of the redshift shell $z=28$ to $z=32$ would contain one such object.
}
\label{fig1}
\end{figure}

Figure 4 shows the density of galaxies versus
the maximum absorption cross section 
(i.e., in the plane
of the large circle centered on the galaxy perpendicular to the line of sight)
with $\delta T<-100$~mK.
Here we point out four fundamental 
and potentially ground-breaking applications
regarding cosmology and galaxy formation,
if a 21-cm tomographic survey of galaxies in the redshift shell
$z=30-40$ is carried out,
which may be able to detect millions of galaxies.

First, a characteristic sharp fall-off at $5-10$ square arcseconds and
a characteristic peak of the number of 21-cm absorption halos is expected,
as seen in Figure 4 due to a lower star formation efficiency in (and low mass of)
minihalos, as suggested by available simulations (Abel \etal 2002).
This should yield direct information
on physics of cooling and star formation in first galaxies,
which may be unobtainable otherwise by any other means in the foreseeable
future.
Note that the left cutoff and the peak density 
are functions of $c_*$ and $g({\rm IMF})$ (noting
the differences between SAL and VMS cases in Figure 4),
where $g({\rm IMF})$ denotes dependence on the properties 
of IMF such as stellar lifetime and spectrum.
A full parameter space exploration will be given in a separate paper with
more detailed treatments.
We expect that $c_*$ and $g({\rm IMF})$ 
may be determined separately,
when jointly analyzed with
the density of absorption halos in the context of the standard CDM model.

Second, we see that the density of strong 21-cm absorption halos
depends strongly on $n_s$,
as testified by the large difference 
(a factor of $\sim 50$) between solid and dotted curves in Figure 4.
One may then obtain a constraint on $n_s$,
which is made possible because the effect due to
difference in IMF may be isolated out, as discussed above,
thanks to the features in the density of absorption halos
(e.g., peak location and sharp fall-off at the low end).
Let us estimate a possible accuracy of such measurements.
At $n\sim 10^{-6}$h$^3$Mpc$^{-3}$ one would find 
0.1 million galaxies in the redshift shell between $z=28$ and $z=32$,
giving a relative fraction (Poisson) error 
of $0.3\%$.
By comparing the solid and dotted curves in Figure 4,
we find that a constraint on $n_s$ 
with $\Delta n_s=0.01$ ($\sim 3\sigma$) may be achieved.
This may have the potential to discriminate between 
inflationary theories (e.g., Liddle \& Lyth 1992; Peiris \etal 2003).
In addition,
the constraint placed on the temperature (or mass) of dark matter particles
or running of the spectral index
may be still tighter, because a significant, finite dark matter temperature 
or a running index tends to 
suppress small-scale power exponentially thus amplify the effects. 
The high-sensitivity constraint on small-scale power 
is afforded by the physical fact that we are dealing with 
rare $\ge 5-6\sigma$ peaks in the matter distribution.

Third, clustering of galaxies may be computed 
using such a survey containing potentially hundred of thousands
to millions of galaxies
in a comoving volume of size $\sim 100$Gpc$^3$ (for the redshift
shell $z=28-32$).
Both the survey volume and the number of observable galaxies within
are large enough to allow for accurate determinations
of the correlations of first galaxies,
particularly on large scales.
It may then provide an independent,
perhaps ``cleaner" characterization of interesting features
in the power spectrum such as the baryonic oscillations,
with the advantage that they are not subject to
subsequent complex physical processes, including
cosmological reionization, gravitational shock heating
of the IGM and complex interplay
between galaxies and IGM, which in turn might 
introduce poorly understood biases in galaxy formation.
A comparison between clustering of first galaxies
and local galaxies (e.g., Eisenstein \etal 2005)
will provide another, high-leverage means to gauge
gravitational growth and other involved processes between $z=30$ to $z=0$.

Finally, the scale ($\sim 1$Mpc comoving) 
of the 21-cm absorption halo signals is
much greater than the nonlinear scale and virial radius (both $\sim 1$kpc comoving).
Thus, the IGM region in the 21-cm absorption halo 
is expected to closely follow the Hubble flow.
Since near \lya\ photons (between \lya\ and Ly$\beta$)
are not subject to absorption by hydrogen (and helium) atoms whose distribution might be complex, 
they escape into the IGM in a spherical fashion.
Additionally, since the dependence on $\Delta$ is linear (see equation 2),
density inhomogeneities are likely to average out (to zero-th order)
and results do not depend sensitively on uncertain linear density fluctuations in the IGM.
Although there might exist ``pores" in the domain of 21-cm absorption halo due to 
fluctuations in local IGM temperatures which may be
caused by local shock heating due to formation of minihalos,
the overall effect is likely negligible,
because the mass fraction contained in all halos down to
a mass as small as $M_h=10^{5}\msun$ is about $10^{-4}$ at $z=30$.
Furthermore, at $n=10^{-6}$h$^3$Mpc$^{-3}$ 
the mean separation between the galaxies
is $100$~Mpc/h, much larger than
the size of \lya\ scattering regions of size $\sim 1$Mpc/h,
so overlapping of the latter should be very rare (taking
into account the known fact that they are strongly clustered
in the standard cosmological model with {\it gaussian} random fluctuations;
Mo \& White 1996).
For these reasons, each 21-cm absorption halo 
is expected to be highly spherical in real space.
Therefore, 21-cm absorption halos
are ideal targets to apply the Alcock-Paczy\'nski (1979) test. 
Accurate measurements of angular size $\Delta\theta$ and radial depth 
$\Delta v$ for a sample of galaxies
would yield a sample of $d_A(z) H(z)$,
where $d_A(z)$ and $H(z)$ 
are the angular diameter distance and Hubble constant, respectively,
both of which are, in general, functions of $\Omega_M$, $w$($\equiv p/\rho$) and $k$,
with $w$ describing the equation of state for dark energy and
$k$ being the curvature of the universe.
As an example, let us assume 
that $\Omega_M$($\approx 0.3$) has been fixed
exactly by independent observations and $k=0$ 
and that $w\approx -1$.
Then one obtains $|d ln[d_A(z) H(z)]/dw|=0.45$ at $z=30$
(Huterer \& Turner 2001).
Let us suppose a relative measurement error
on each individual $d_A(z) H(z)$ is $20\%$,
then with ten thousand galaxies,
one could obtain a highly accurate
constraint on $w$ with $\Delta w = 20\%/0.45/\sqrt{10000} \sim 0.004$.
Likely, the accuracy of $w$ determined by this method may eventually be limited by 
the accuracy with which $\Omega_M$ (and $k$) can be determined
by independent observations, due to the degenerate nature.
We stress that this method 
is valid for each individual first galaxy and unaffected
by uncertainties, for example,
in the precise abundance of such galaxies.

In post-survey analyzes 
one faces the practical issue of extracting the wanted
signals from the raw data, whose amplitude is expected to be dominated by 
foreground radio sources, including galactic synchrotron radiation,
galactic and extragalactic free-free emission, and extragalactic point sources
(e.g., Di Matteo, Ciardi, \& Miniati 2004).
While seemingly daunting,
it has already been shown that signals of the amplitude  
proposed here may be recovered with relatively high fidelity,
when one takes into account the expected, potent differences
in the spectral and angular properties between 
the 21-cm signal and foreground contaminants 
(e.g., Zaldarriaga, Furlanetto, Hernquist 2004; 
Santos, Cooray, \& Knox 2005;
Wang \etal 2005). 
Since the 21-cm absorption halos are expected to be rather regular and simple,
one might be able to significantly enhance the signal by
using additional techniques, such as matched filter algorithm,
in combination with foreground ``cleaning" methods.
Finally, 
the amount of data in such a high spatial and frequency resolution
3-dimensional survey will be many orders of magnitude larger than that of WMAP.
Computational challenges for analyzing it
will be of paramount concern and most likely demand new
and innovative approaches.

\section{Conclusions}

It is shown that a first galaxy 
hosted by a halo of mass $M=10^{7.5}-10^9\msun$ at $z=30-40$ 
possesses a large 21-cm absorption halo against the CMB
with a brightness temperature decrement $\delta T=-(100-150)$~mK 
and an angular size of $10^{''}-100^{''}$.
A 21-cm tomographic survey of galaxies in the redshift shell
at $z=30-40$ may detect millions of galaxies 
and may yield critical information on cosmology and galaxy formation.
A successful observation may need
an angular resolution of $\le 1^{''}$,
a spectral resolution of $\le 4$kHz,
and a sensitivity of $\le 10$~mK at $35-45$~MHz.
LOFAR appears poised to be able to 
execute this unprecedented task, at least for the high end
of the distribution.

At least four fundamental applications may be launched with such a survey,
which could potentially revolutionize cosmological study 
and perhaps the field of astro-particle physics.
First, it may provide unprecedented
constraint on star formation physics in first galaxies,
for there is a proprietary sharp feature 
related to the threshold halo mass for efficient atomic cooling.
Second, it may provide a unique and sensitive probe
of the small-scale power in the cosmological model 
hence physics of dark matter and inflation,
by being able to, for example,
constrain $n_s$ to an accuracy of $\Delta n_s=0.01$
at a high confidence level.
Constraints on the nature of dark matter particles,
i.e., mass or temperature, or running of index could be still tighter.
Third, clustering of galaxies that may be computed 
with such a survey will provide an independent set
of characterizations of potentially interesting features
on large scales in the power spectrum including the baryonic oscillations,
which may be compared to local measurements
(Eisenstein \etal 2005) to 
shed light on gravitational growth 
and other involved processes from $z=30$ to $z=0$.
Finally, the 21-cm absorption halos
are expected to be highly spherical 
and trace the Hubble flow faithfully,
and thus are ideal systems for an application of the Alcock-Paczy\'nski test.
Exceedingly accurate determinations of 
key cosmological parameters, in particular, 
the equation of state of the dark energy,
may be finally realized.
As an example, it does not seem excessively difficult
to determine $w$ to an accuracy of
$\Delta w\sim 0.01$, if $\Omega_M$ has been determined
to a high accuracy by different means.
If achieved, 
it may have profound ramifications pertaining dark energy and fundamental
particle physics (e.g., Upadhye, Ishak, \& Steinhardt 2005).

If a null detection of the proposed signal is found,
as it might turn out,
implications may be as profound.
It might be indicative of some heating and/or reionizing sources
in the early universe ($z=30-200$) 
that precede or are largely unrelated to structure formation,
possibly due to yet unknown properties of dark matter particles or dark energy.
Alternatively, star formation and/or BH accretion in first galaxies 
may be markedly different from our current expectations.

\acknowledgments
I thank Dr. Daniel Schaerer for helpful information on Pop III stars.
This research is supported in part by grants 
AST-0206299, AST-0407176 and NAG5-13381.

\end{document}